\documentclass[conference]{IEEEtran}
\IEEEoverridecommandlockouts

\usepackage{xcolor}
\usepackage{placeins}
 \usepackage{amsmath,amssymb,amsfonts}
 \usepackage{makecell}
\usepackage{subcaption}
\usepackage{multirow}
\usepackage{graphicx}
\usepackage[space]{grffile}
\usepackage{mathtools, nccmath}
\usepackage{dblfloatfix}
\usepackage{amsmath}
\usepackage{dsfont}
\usepackage{wrapfig}
\usepackage{algorithm}
\usepackage{float}
\usepackage{algpseudocode}
\usepackage{algcompatible}
\usepackage{mathtools}
\usepackage{algorithmicx}
\usepackage{pifont}%
\usepackage{makecell}
\usepackage{todonotes}
\newcommand{\RomanNumeralCaps}[1]
    {\MakeUppercase{\romannumeral #1}}

\def\BibTeX{{\rm B\kern-.05em{\sc i\kern-.025em b}\kern-.08em
    T\kern-.1667em\lower.7ex\hbox{E}\kern-.125emX}}
\begin{document}

\title{Artificial Intelligence (AI)Centric Management of Resources in Modern Distributed Computing Systems}

\author{\IEEEauthorblockN{Shashikant Ilager\textsuperscript{1}, Rajeev Muralidhar\textsuperscript{1,2}, Rajkumar Buyya\textsuperscript{1}}
\IEEEauthorblockA{\textit{\textsuperscript{1}Cloud Computing and Distributed Systems (CLOUDS) Laboratory} \\
\textit{School of Computing and Information Systems, 
The University of Melbourne, Australia}\\
\textsuperscript{2}Amazon Web Services (AWS), Australia\\
}
}

\maketitle
\begin{abstract}

Contemporary Distributed Computing Systems (DCS) such as Cloud Data Centres are large scale, complex, heterogeneous, and distributed across multiple networks and geographical boundaries. On the other hand, the Internet of Things (IoT)-driven applications are producing a huge amount of data that requires real-time processing and fast response. Managing these resources efficiently to provide reliable services to end-users or applications is a challenging task. The existing Resource Management Systems (RMS) rely on either static or heuristic solutions inadequate for such composite and dynamic systems. The advent of Artificial Intelligence (AI) due to data availability and processing capabilities manifested into possibilities of exploring data-driven solutions in RMS tasks that are adaptive, accurate, and efficient. In this regard, this paper aims to draw the motivations and necessities for data-driven solutions in resource management. It identifies the challenges associated with it and outlines the potential future research directions detailing where and how to apply the data-driven techniques in the different RMS tasks. Finally, it provides a conceptual data-driven RMS model for DCS and presents the two real-time use cases (GPU frequency scaling and data centre resource management from Google Cloud and Microsoft Azure) demonstrating  AI-centric approaches' feasibility.
\end{abstract}

\begin{IEEEkeywords}
Distributed Computing, Resource Management, AI Techniques, Edge Computing, Cloud Computing
\end{IEEEkeywords}

\section{Introduction}
Internet-based Distributed Computing Systems (DCS) have become an essential backbone of the modern digital economy, society, and industrial operations. The emergence of the Internet of Things (IoT), diverse mobile applications, smart grids, smart industries, and smart cities has resulted in massive amounts of data generation. Thus, it has increased the demand for computing resources \cite{gubbi2013internet} to process this data and derive valuable insights for users and businesses. According to the report from Norton \cite{norton2019usa},  21 billion IoT devices will be connected to the internet by 2025, creating substantial economic opportunities. Computing models such as Cloud and Edge computing have revolutionised the way services are delivered and consumed by providing flexible on-demand access to services with a pay-as-you-go model. Besides,  new application and execution models like micro-services and serverless or Function as Service (FaaS) computing \cite{baldini2017serverless} are becoming mainstream that significantly reduces the complexities in the design and deployment of software components.
On the other hand, this increased connectivity and heterogeneous workloads demand distinct Quality of Service (QoS) levels to satisfy their application requirements\cite{gan2019open, dastjerdi2016fog, fox2009above}. These developments have led to building hyper-scale data centres and complex multi-tier computing infrastructures that require new innovative approaches in managing resources efficiently and provide reliable services. Deployment of 5G and related infrastructures like dynamic network slicing for high bandwidth, high throughput, and low latency applications has only increased the challenges.

Resource Management Systems (RMS) in DCS's are middleware platforms that perform different tasks such as resource provisioning, monitoring, workload scheduling, and many others. Building an efficient RMS for the present and imminent distributed systems are challenging due to many reasons. Significantly, the new class of applications, networks, and Cyber-Physical-Systems (CPS) such as data centres are enormously complex and challenging to fine-tune their parameters manually. For example, "Just $10$ pieces of equipment, each with $10$ settings, would have $10$ to the $10^{th}$ power, or $10$ billion, possible configurations — a set of possibilities far beyond the ability of anyone to test for real" \cite{schwartz2019allen, amodei2018ai}. The emerging network technologies, including 5G and satellite networks, such as Amazon's Project Kuiper and SpaceX's StarLink, have opened up new dimensions \cite{giambene2018satellite} and opportunities for developing advanced applications that require high bandwidth, high availability, and low latency. The availability of massive data and advancement in computing capabilities has witnessed the resurgence of Artificial intelligence (AI) techniques driving innovation across different domains such as healthcare, autonomous driving, and robotics \cite{giambene2018satellite, russell2002artificial}. Training AI models itself consumes vast resources and is increasing exponentially and doubling every 3.4 months for the largest AI models (compared to Moores' Law' 2-year doubling period) \cite{schwartz2019allen}. The Cloud and Edge infrastructures deliver resources  (compute, network, storage)  required to accommodate these rapid changes across different domains managed by third-party service providers. These are highly distributed, large-scale, and contain numerous heterogeneous resources. Furthermore, they are multi-tenant, with users sharing the underlying resources with diverse workload characteristics. Thus, providing the performance requirements in such a shared environment and increasing resource utilisation is a critical problem and challenging for RMS \cite{buyya2018manifesto}.

The existing RMS techniques from operating systems to large scale DCS's are predominantly designed and built using preset threshold-based rules or heuristics. These solutions are static and often employ reactive solutions \cite{bianchini2020toward}; they work well in the general case but cannot adjust to the dynamic contexts \cite{dean2017machine}. Moreover, once deployed, they considerably fail to adapt and improve themselves in the runtime. In complex dynamic environments (such as Cloud and Edge), they are incapable of capturing the infrastructure and workload complexities and hence fall through. Consequently, the AI-centric approaches built on actual data and measurements collected from respective DCS environments are more promising, perform better, and adapt to dynamic contexts. Unlike heuristics, these are data-driven models built based on historical data. Accordingly, AI-centric methods can employ proactive measures by foreseeing the potential outcome based on current conditions. For instance, a static heuristic solution for scaling the resource uses workload and system load parameters to trigger the scaling mechanism. However, this reactive scaling diminishes the users' experience for a certain period (due to the time required for system bootup and application trigger).
Consequently, an AI-centric RMS enabled by data-driven Machine Learning (ML) model can predict the future workload demand and scale up or scale down the resources beforehand as needed. Such techniques are highly valuable for both users to obtain better QoS and service providers to offer reliable services and retain their business competency in the market. Moreover, methods like Reinforcement Learning (RL) \cite{ dean2017machine, sutton2018reinforcement} can improve RMS's decisions and policies by using monitoring and feedback data in runtime, responding to the current demand, workload, and underlying system status.

AI-centric RMS in DCS is more feasible now than ever for multiple reasons. First, AI techniques have matured and have proven efficient in many critical domains such as computer vision, natural language processing, healthcare applications, and autonomous vehicles. Second, most DCS platforms generate enormous amounts of data currently pushed as logs for debugging purposes or failure-cause explorations. For example, Cyber-Physical-Systems (CPS) in data centres already have hundreds of onboard CPU and external sensors monitoring workload, energy, temperature, and weather parameters. Such data is useful to build ML models cost-effectively. Finally, the increasing scale in computing infrastructure and its complexities require automated resource management systems that can deliver the decisions based on the data and key-insights from experience, to which AI models are ideal.

In this regard, this paper makes the following key contributions: (1) presents evolution, and the state-of-the-art RMS techniques in DCS, (2) enlists the challenges associated with data-driven RMS methods, (3) identifies the future research directions and point out the different tasks in which AI-centric methods can be efficiently applied and benefited from, (4) proposes a conceptual data-driven RMS model, and (5) demonstrates two real-time use-cases using data-driven AI methods (related to energy-efficient GPU clock configurations and management of resources in data centres).

The rest of the paper is organised as follows. Section \RomanNumeralCaps{2} gives an overview of DCS evolution and state-of-the-art practices in RMS. Section \RomanNumeralCaps{3} identifies the challenges associated with data-driven methods. Section \RomanNumeralCaps{4} draws Future research directions. In Section \RomanNumeralCaps{5}, a conceptual AI-centric RMS model is presented, and Section \RomanNumeralCaps{6} demonstrates the feasibility of AI-centric methods using two real-time use cases. Finally, the conclusion is drawn in Section \RomanNumeralCaps{7}.
\section{DCS Evolution and the State-of-the-Art}

\begin{figure*}
\captionsetup{justification=centering}
\includegraphics[width=\linewidth]{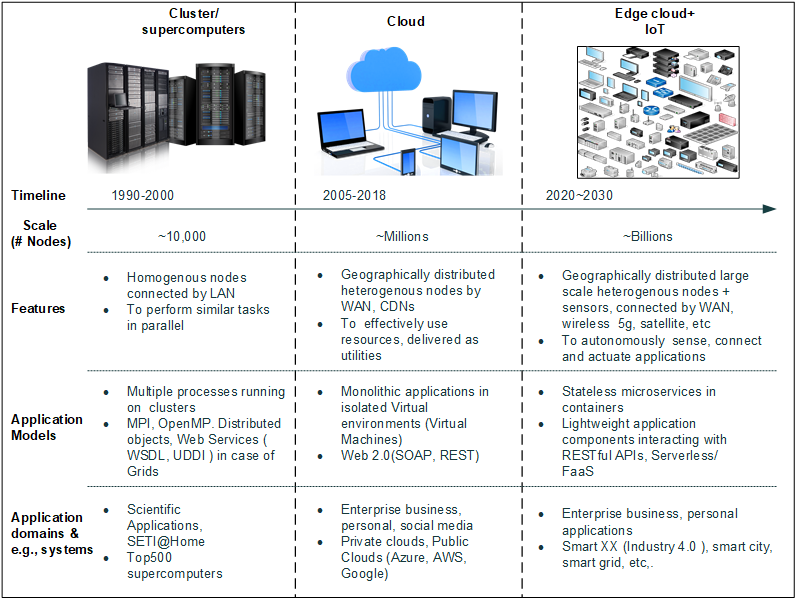}
\caption{An overview of contemporary DCS evolution (Timeline shows approximate time of the genesis of the system and evolved as mainstream with some overlapping’s. The points shown for all dimensions are representative but not exhaustive and only lists the important facets.)}
\label{fig:overview}
\end{figure*}

An overview of the evolution of primary DCS's is given in Figure \ref{fig:overview}. Early DCS systems are prominently used in scientific domain applications composed of parallel tasks (distributed jobs in grid computing) and executed on clusters or supercomputing systems. The development of technologies such as service-orientated computing (Web services, REST, SOAP, etc.), virtualisation, and demand for utility-oriented services created the current Cloud computing-based data centres. However, the next decade of DCS's will be driven by IoT-based applications and scenarios that need to process the enormous amount of data and derive meaningful intelligence and business values from it. These IoT-based applications consist of numerous sensors and computing nodes distributed across different network layers from  Edge to remote Cloud. Thus, requiring an autonomic sense-connect-actuate model \cite{gubbi2013internet} where application tasks are composed, deployed, and executed autonomously—demanding additional machine-to-machine interactions (compared to the current human-to-machine interactions). RMS should autonomously provision resources, schedule application tasks, and manage their demand for QoS and low latency.

In parallel to system advancements, application models have continued to evolve and create new software design patterns like micro-services and execution models like serverless or Function as Service (FaaS) computing. To that end, managing these modern resources and applications requires intelligent decisions enabled from the AI-centric solutions. Although AI-centric RMS techniques will be applicable for all the computing paradigms discussed here, we mainly keep our discussions and illustrations around the Cloud and Edge computing paradigms.

With the increased scale and complexities in next-generation DCSs, traditional static or heuristics solutions are becoming inadequate. These methods require careful hand pruning and human intervention to adapt to the dynamic environments \cite{dean2017machine}. Consequently, AI-centric data-driven solutions are promising, and there have been many attempts in recent years to address the resource management problems using the data-driven ML solutions \cite{bianchini2020toward}. For example, Google has achieved a 40\% efficiency in managing its cooling infrastructure using simple ML techniques and learning from historical data \cite{gao2014machine}. Many other methods explored problems such as device placement, scheduling, and application scaling using data-driven methods \cite{mirhoseini2017device}, \cite{tulifog}. At the system architecture level, \cite{ayers2019asmdb} used massive data sets of hardware performance counters and profiles collected from large-scale  Google data centre servers and utilised this data to reason, analyse and mitigate front-end stalls in warehouse-scale systems.
However, data-driven AI solutions for RMS are in its superficial stage. They require meticulous attention to address the challenges they pose and simultaneously identify potential avenues to incorporate these methods. Moreover, it is essential to build the general frameworks and standards to adopt AI solutions in resource management that are scalable and manageable.

\section{Challenges }\label{Section:challenges}
In this section, we identify and describe the critical issues associated with the adoption of AI solutions in the resource management of distributed computing systems.

\subsection{Availability of Data}
The quality of data used to train the models determines the success of machine learning techniques. Also, this data should be available in large quantities with enough features covering all the aspects of environments \cite{cummins2017end, cano2019optimizing}. Within DCS, multiple challenges exist concerning the availability of such data. First, currently, different resource abstraction platforms collect the data at different granularity. The physical machine-level data from onboard sensors and counters is gathered and accessed by tools like Intelligent Platform Management Interface (IPMI), while at a higher abstraction level, middleware platforms collect data related to workload level, user information, and surrounding environmental conditions (temperature, cooling energy in the data centre). Also, network elements such as SDN controllers collect data related to network load, traffic, and routing. Unifying these data together and preprocessing it in a meaningful way is a complex and tedious task. The respective tools gather the data in a different format without common standards between them. Hence, building data-pipelines combining various subsystems data is crucial for the flexible adoption of ML solutions. Secondly, current monitoring systems collect data and push them into logging repositories to be used later for debugging. However, converting this data for ML-ready requires monotonous data-engineering. Hence, future systems should be explicitly designed to gather information that can be directly fed to the ML models with minimal data engineering and preprocessing effort. Lastly, although several publicly available datasets provide workload traces, there are hardly any public datasets available representing various infrastructure, including physical resource configurations, energy footprints, and several other essential parameters (due to privacy and NDAs). Therefore, getting access to such data is a challenge and needs collaborative efforts and data management standards from the relevant stakeholders. Moreover, requiring standardised data formats and domain-specific frameworks \cite{portugal2016survey}.

\subsection{Managing the Deployment of Models}\label{sec:managedeploy}
Training ML models and inference in runtime needs an expensive amount of computational resources. However, one significant challenge is to manage the life cycle of ML models, including deciding how much to train, where to deploy the training modules in multi-tier computing architectures like Edge/Fog. As resources have limited capabilities at a lower level and should be allocated to needful applications, if these scarce resources are predominantly used to train models or run the RL agents, the latency-sensitive applications will experience resource starvation. On the other hand, if the models (RL agents) are trained or deployed in resource enriched cloud, the latency to push the inference decisions or the runtime feedback data to edge nodes shoots up, thus creating a delay-bottlenecks in RMS decisions. Furthermore, ML models tend to learn too much with the expense of massive computational resources. Therefore, the innovative solutions are needed to decide how much learning is sufficient based on specific constraints (resource budget, time-budget, etc.) and estimate context-aware adaptive accuracy thresholds of ML models \cite{toma2019adaptive}. To overcome this, techniques like transfer learning, distributed learning can be applied to reduce computational demands \cite{cano2019optimizing}. In addition, dedicated CPUs, GPUs, and domain-specific accelerators like Google TPU, Intel Habana, and FPGAs (Azure)  can carry out the inference.

\subsection{Non-Deterministic Outputs }
Unlike statistical models, which are analogous for its deterministic outputs, ML models are intrinsically exploratory and depend on stochasticity for many of its operations, thus producing the non-deterministic results. For example, the cognitive neural nets, which are basic building blocks for many regressions, classification, and Deep Learning (DL) algorithms primarily rely on the principles of stochasticity for different operations (stochastic gradient descent, exploration phase in RL). When run multiple times with the same inputs, they tend to approximate the results and produce different outputs \cite{russell2002artificial}. This may pose a severe challenge in the DCS, such as Edge and Clouds, where strict Service Level Agreements (SLAs) govern the delivery of services requiring deterministic results. For example, if a service provider fixes a price based on certain conditions using ML models, consumers expect the price to be similar in all the time under similar settings. However, ML models may have a deviation in pricing due to stochasticity creating the transparency issues between users and service providers. Many recent works have focused on this issue and introduced techniques such as induced constraints in neural nets to produce the deterministic outputs \cite{lee2019gradient}. Yet, stochasticity in the ML model is inherent and requires careful monitoring and control over its output.

\subsection{Black Box Decision Making }
The ML models' decision-making process follows a completely black-box approach and fails to provide satisfactory justification for its decisions. The inherent probabilistic architectures and enormous complexities within ML models make it hard to evade the black-box decisions. It becomes more crucial in an environment such as DCS, where users expect useful feedback and explanation for any action taken by the service provider. This is instrumental in building trust between service providers and consumers. For instance, in case of a high overload condition, it is usual that service provider shall preempt few resources from certain users with the expense of certain SLA violations. However, choosing which users' resources should be preempted is crucial in business-driven environments. This requires simultaneously providing fair decisions and valid reasons. Many works have undertaken to build the explanatory ML models (Explainable AI- XAI) to address this issue \cite{arrieta2020explainable, gunning2017explainable}. However, solving this continues to remain a challenging task.
\subsection{Lightweight and Meaningful Semantics}
The DCS environment having heterogeneous resources across the multi-tiers accommodates different application services.  RMS should interact between different resources, entities, and application services to efficiently manage the resources. However, these requisites semantic models that represent all these various entities meaningfully. Existing semantic models are either heavy or inadequate for such complex environments. Therefore, lightweight semantic models are needed to represent the resource, entities, applications, and services without introducing the overhead \cite{bermudez2016iot}.

\subsection{Complex Network Architectures, Overlays, Upcoming Features}
Network architectures across DCS and telecom networks are evolving rapidly using software-defined infrastructure, hierarchical overlay networks, Network Function Virtualization (NFV), and Virtual Network Functions (VNF). Commercial clouds like Amazon, Google, and Microsoft have recently partnered with telecom operators worldwide to deploy ultra-low latency infrastructure (AWS Wavelength and Azure Edge Zone, for example) for emerging 5G networks. The explosion of data from these 5G deployments and resource provisioning for high bandwidth, throughput, and low latency response through dynamic network slicing requires a complex orchestration of network functions \cite{zhang2017networkslice}.

In future DCS, RMS needs to consider these complex network architectures, the overlap between telecom and public/private clouds, and service function orchestration to meet end-to-end bandwidth, throughput, and latency requirements. These architectures and implementations, in turn, generate enormous amounts of data at different levels of the hierarchical network architecture. As different types of data are generated in different abstraction levels, standardised well-agreed upon data formats and models for each aspect needs to be developed.

\subsection{Performance, Efficiency, and Domain Expertise}
Many ML algorithms and RL algorithms face performance issues like a cold-start problem. Specifically, RL algorithms spend a vast amount of the initial phase in exploration before reaching its optimal policies creating an inefficient period where the decisions are suboptimal, even completely random or incorrect leading to massive SLA violations \cite{cano2019optimizing}. RL-based approaches also face several challenges in the real world including (1) need for learning on the real system from limited samples, (2) safety constraints that should never or at least rarely be violated, (3) need of reward functions that are unspecified, multi-objective, or risk-sensitive, (4) inference that must happen in real-time at the control frequency of the system \cite{dulac2019challenges}. In addition, AI models are compute-heavy and designed with a primary focus on accuracy-optimisation resulting in a massive amount of energy consumption \cite{schwartz2019allen}. Consequently, new approaches are needed to balance the trade-offs between accuracy, energy, and performance overhead.
Furthermore, current ML algorithms, including neural network architectures/libraries, are primarily designed to solve computer vision problems. Adapting them to RMS tasks needs some degree of transformation of the way input and outputs are interpreted. Currently, many AI-centric RMS algorithms transform their problem space and further use simple heuristics to interpret the result back and apply to the RMS problems. Such complexities demand expertise from many related domains. Thus, newer approaches, algorithms, standardised frameworks, and domain-specific AI frameworks are required to adopt AI in RMS efficiently.

\section{Future Research Directions}
Despite the challenges associated, AI solutions provide many opportunities to incorporate these techniques into RMS and benefit from them. In this section, we explore different avenues where AI techniques can be applied to manage distributed systems resources.

\subsection{Data-driven Resource Provisioning and Scheduling}
Resource provisioning and scheduling are a fundamental element of an RMS. Usually, resources are virtualised, and specifically, computing resources are delivered as Virtual machine (VM) or lightweight containers. The problems related to provisioning such as estimating the number of resources required for an application, co-locating workloads based on their resource consumption behaviours and several others can be addressed using AI techniques. These techniques can be extended to special case provisions such as spot instances. Utilising spot instances for application execution needs careful estimation of application run time (to avoid the state corruption or loss of computation if resources are preempted) and accordingly deciding resource quantity and checkpointing logic. It may require building prediction models based on previous execution performance counters or correlating with clusters based on existing knowledge base \cite{shashiccgrid2020}.

In edge computing environments, RMS should utilise resources from multi-tier infrastructure, and selecting nodes from different layers also requires intelligence and adaptation to application demands and infrastructure status. Furthermore, data-driven AI solutions can be used in scheduling tasks such as finding an efficient node, VM consolidation, migration, etc. The prediction models' historical data and adaptive RL models can be used to manage dynamic scheduling and resource provisioning.

\subsection{Managing Elasticity using Predictive Analytics}

Elasticity is an essential feature providing flexibility by scaling up or scaling down the resources based on the applications' QoS requirements and budget constraints. Current approaches in elasticity are based on the reactive methods where resources are scaled according to the system load (in terms of the number of users and input requests). However, such reactive measures diminish the SLAs due to bootup time and swift burst loads. In contrast, forecasting the future load based on the application's past usage behaviours and proactively scaling the resources beforehand vastly improves SLAs and saves costs. Essentially, it needs time series analysis to predict future load using methods like ARIMA or more advanced RNN techniques such as LSTM networks that are proven to be efficient in capturing the temporal behaviours \cite{gan2019leveraging}. Such proactive measures from service providers enable efficient management of demand response without compromising the SLAs.

\subsection{Energy Efficiency and Carbon footprint Management}
One of the major challenges of computing in recent years has been energy consumption. Increasing reliance on computing resources has created enormous energy, economic and environmental issues. It is estimated that by 2025, data centres itself would consume around 20\% of global electricity and emit up to 5\% of the world's carbon emissions \cite{Lima2017}. Energy efficiency can be achieved across the computing stack from managing hardware circuits to data centre level workload management. Recent studies have shown promising results of AI techniques in device energy-optimised frequency management \cite{shashiccgrid2020}, intelligent and energy-efficient workload management (scheduling, consolidation), reducing cooling energy by fine-tuning cooling parameters \cite{gao2014machine, ilager2019etas}, and executing applications within power budgets \cite{bianchini2020toward}, etc. In addition, it can also be effectively used in minimising the carbon-footprints by forecasting renewable energy and shifting the workloads across clouds accordingly. Each of these subproblems can be addressed by using a combination of predictive and RL models based on application scenarios and requirements.

\subsection{Security and Privacy Management}
As cybersystems have become sophisticated and widely interconnected, preserving the privacy of data and securing resources from external threats has become quintessential. Dealing with security has the implications far beyond resource management, including privacy-preserving and complying with the respective jurisdiction's rules. For instance,  RMS with user-level schedulers can classify input records and process records with privacy sensitivity within local resource environments (e.g., private cloud) and others on public clouds. One such work is carried out by the University of Washington \cite{XUanekasecuroty} wherein a deep learning method is used to classify medical records into sensitive and nonsensitive based on data privacy. They created a user-level scheduler for the Aneka Cloud application platform and able to process sensitive medical records on their private cloud and nonsensitive records on Amazon AWS EC2 public cloud.  

If resources are maliciously compromised, RMS should adapt to the requirements of the security concerns. There has been widespread use of ML algorithms in many aspects of security management. It includes AI-based Intruder Detection Systems (IDS) to prevent unauthorized access, anomaly detection \cite{moghaddam2019acas,butun2015anomly} to identify the deviations in the application/ resource behaviors. AI techniques, including Artificial Neural Networks (ANNs), ensemble learning, Bayesian networks, association rules, and several classification techniques like SVM, can be effectively utilised to address these security-related problems \cite{buczak2015survey}. They can also be predominantly used in preventing Denial-of-service attacks (DDoS) by analysing traffic patterns and filtering suspected traffic, hence, preventing the system failures \cite{yuan2017deepdefense}.  Such measures vastly help to manage the resources securely and thus increasing the reliability of the system.

\subsection{Managing Cloud Economics}
Cloud economics is a complex problem and requires vast domain knowledge and expertise to price services adequately. It is also essential for consumers to easily understand pricing models and estimate the cost for their deployments. Current pricing models largely depend on subscription types, e.g., reserved, on-demand, or spot instances. The pricing for these subscription models is driven by standard economic principles like auction mechanisms, cost-benefit analysis, profit, revenue maximisation, etc. These pricing problems are solved using techniques like Operation Research (OR) or stochastic game theory approaches \cite{mistry2018economic}. However, such methods are mostly inflexible, and they either overprice the services or results in loss of revenues for cloud service providers. In this regard. ML models can forecast resource demand, and accordingly, excessive resources can be pooled in the open market for consumers. In addition, pricing can be more dynamic based on this forecasted demand response that benefits both consumers and service providers.
\subsection{Generating the Large-scale Data Sets}
Machine learning models require large amounts of training data for improved accuracy. However, access to large scale data is limited due to privacy and lack of capabilities to generate a large quantity of data from infrastructure. To that end, AI models itself can be used to create large-scale synthetic datasets that closely depict the real-world datasets. For instance, given a small quantity of data as input, Generative Adversarial Networks (GANs) can be used to produce large-scale data \cite{zhang2018generative}. Such methods are highly feasible in generating time-series data of DCS infrastructure. Moreover, these methods can also be leveraged to produce datasets from the incomplete datasets adequately. Such large-scale data sets are necessary to train efficient predictive models and bootstrap the RL agents to achieve a reasonable efficiency in its policies.

\subsection{Future System Architectures}
Cloud services have recently undergone a shift from monolithic applications to microservices, with hundreds or thousands of loosely-coupled microservices comprising the end-to-end application. In \cite{gan2019open}, the authors explore the implications of these microservices on hardware and system architectures, bottlenecks therein, and lessons for future data centre server design. Microservices affect the computation to communication ratio, as communication dominates, and the amount of computation per microservice decreases. Similarly, microservices require revisiting whether big or small servers are preferable. In \cite{ayers2019asmdb}, the authors use an always-on, fleet-wide monitoring system, to track front-end stalls, I-cache and D-cache miss (as cloud microservices do not lend them amenable to cache locality unlike traditional workloads) across hundreds and thousands of servers across Google's warehouse-scale computers. The enormous amounts of data generated and analysed help provide valuable feedback for the design of next-generation servers. Similarly, deep learning can be used to diagnose unpredictable performance in cloud systems. Data from such systems can thus be invaluable for the hardware and system architectures of future DCS.

\subsection{Other Avenues}
Along with the aforementioned directions, AI-centric solutions can be applied to several other RMS tasks, including optimising the heuristics itself \cite{cummins2017end}, network optimisations (e.g., TCP window size, SDN routing optimisation problems), and storage infrastructure management \cite{cano2019optimizing}. Moreover, learning-based systems can be extended across different computing system stack, from lower abstraction levels, including hardware design, compiler optimisations, operating system policies, to a higher level interconnected distributed system platforms\cite{dean2017machine}.
\section{Conceptual Model for AI-centric RMS}

\begin{figure}
\captionsetup{justification=centering}
\includegraphics[width=\linewidth]{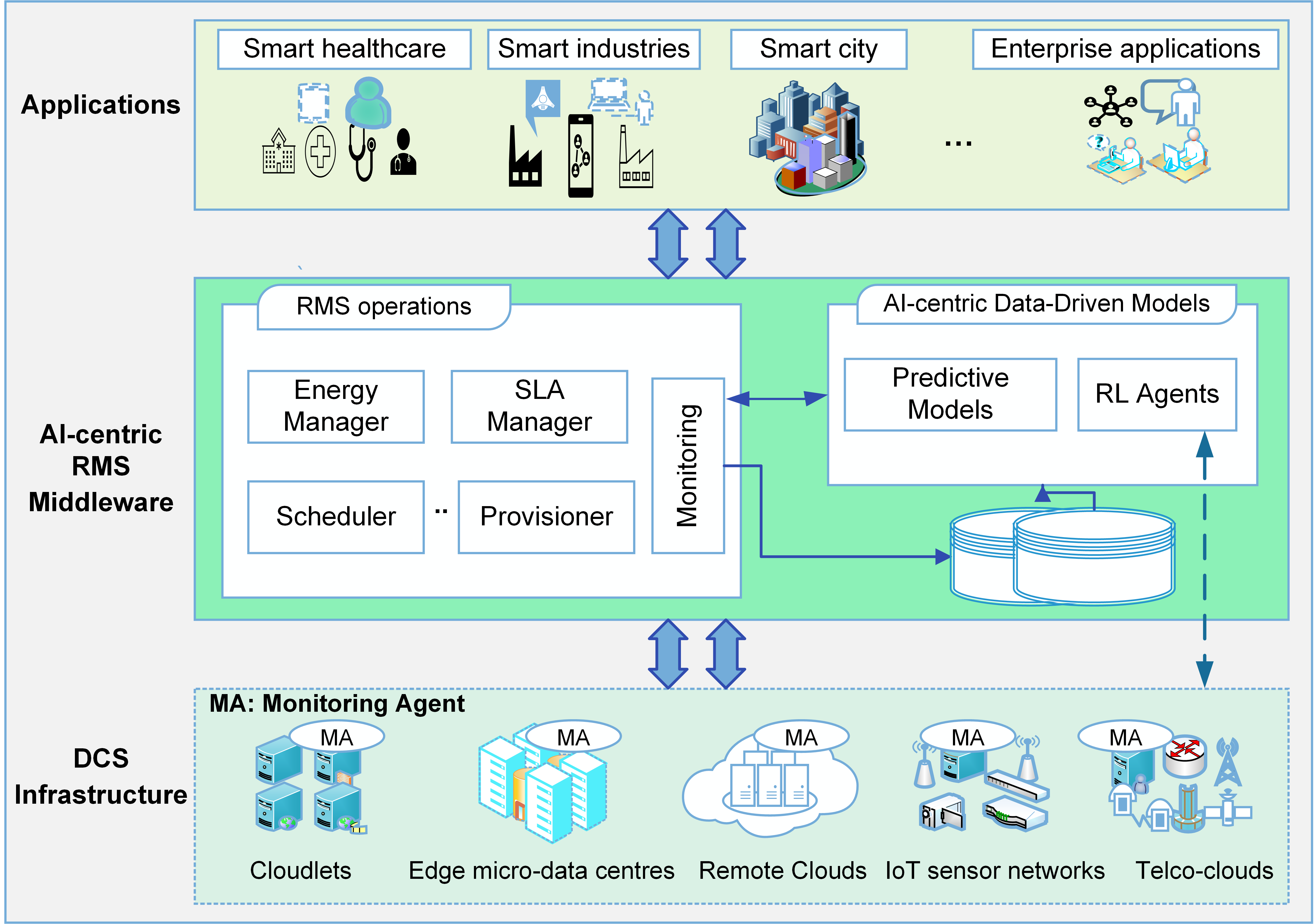}
\caption{Conceptual Data-Driven RMS Model}
\label{fig:conceptual_model}
\end{figure}

In the AI-centric RMS (Resource Management Systems) system, models need to be trained and deployed for the inference used by the RMS for different tasks. However, integrating data-driven models into DCS platforms in a scalable and generic manner is challenging and is still at a conception stage. In this regard, as shown in Figure \ref{fig:conceptual_model}, we provide a high-level architectural model for such data-driven RMS. The essential elements of this system are explained below. It consists of three entities:
\\
\textbf{Users/ Applications:} Users requiring computing resources or services interact with the middleware using APIs or interfaces.
\\
\textbf{AI-centric RMS Middleware:} This is responsible for performing different tasks related to managing user requests and underlying infrastructure. The AI-centric RMS tasks continuously interact with the data-driven models for accurate and efficient decisions. The RMS needs to perform various tasks, including provisioning the resources, scheduling them on appropriate nodes, monitoring in runtime, dynamic optimisations like migrations, and consolidations \cite{bianchini2020toward} to avoid the potential SLA violations. Traditionally, these tasks are done using the algorithms implemented within the RMS system that would execute the policies based on the heuristics or threshold-based policies. However, in this AI-centric RMS, the individual RMS operations are aided with inputs from the data-driven models. The data-driven AI models are broadly categorised into two types, (1) predictive models and (2) adaptive RL models. In the former, models are trained offline using supervised or unsupervised ML algorithms utilising historical data collected from the DCS environment that includes features from resources, entities, and application services. This data is stored in databases, and data-engineering is done, such as preprocessing, cleaning, normalising, to suit AI models' requirements. Thus, this offline training can be done on remote cloud nodes to benefit from the specialised, powerful computing resources. The trained models can be deployed on specialised inference devices like Google Edge TPU and Intel Habana. Choosing the optimal place and deciding where to deploy these ML models depends on where the RMS engine is deployed in the environment, and this is itself a challenging research topic that should be addressed as described in Section \ref{sec:managedeploy}. 
In the latter case, runtime adaptive models such as Reinforcement Learning (RL) that continue to improve their policies based on agents' interactions and system feedback. It requires both initial learning and runtime policy improvement methods that need to be updated after every episode (certain time reaching to terminal state). The RMS operations can interact with both the predictive and RL-based data-driven models using the RESTful APIs in runtime \cite{bianchini2020toward}.
\\
\textbf{DCS Infrastructure:} The computing infrastructure comprises heterogeneous resources, including sensors, gateway servers, edge data centres, and remote clouds.
Therefore, adopting the data-driven AI-centric RMS models needs a significant change in the way current RMS systems are designed and implemented, as well as monitoring agents, interfaces, and deployment policies that can be easily integrated into existing environments.
\section{Demonstration Case Studies}
In this section, we present two use cases that have applied ML techniques to the following problems: (1) data-driven configuration of device frequencies for energy-efficient workload scheduling in cloud GPUs\cite{shashiccgrid2020}, (2) data centre resource management using ML models \cite{bianchini2020toward, gao2014machine}.
\subsection{Data-Driven GPU Clock Configuration and Deadline-aware Scheduling}
\begin{figure}
\captionsetup{justification=centering}
\includegraphics[width=\linewidth]{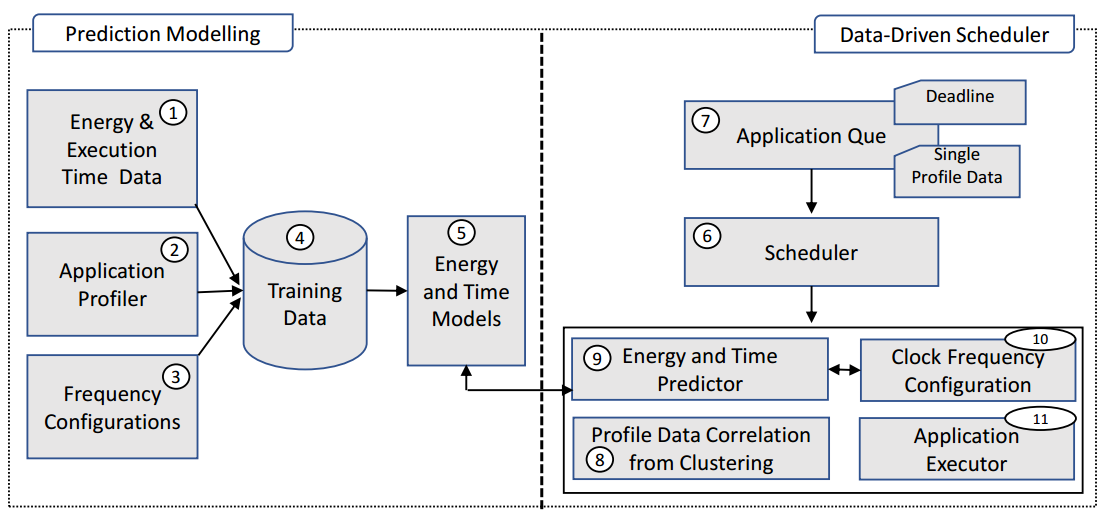}
\caption{System Model}
\label{fig:system_model}
\end{figure}

Graphics Processing Units (GPUs) have become the de-facto computing platform for advanced compute-intensive applications such as video processing and autonomous cars. Additionally, ML models are massively reliant on the GPUs for training due to their efficient SIMD architectures that are highly suitable for parallel computations. However, the energy consumption of GPUs is a critical problem. Dynamic Voltage Frequency Scaling (DVFS) is a widely used technique to reduce the dynamic power of GPUs. Yet, configuring the optimal clock frequency for essential performance requirements is a non-trivial task due to the complex nonlinear relationship between the application's runtime performance characteristics, energy, and execution time. It becomes even more challenging when different applications behave distinctively with similar clock settings. Simple analytical solutions and standard GPU frequency scaling heuristics fail to capture these intricacies and scale the frequencies appropriately. In this regard, we propose a data-driven frequency scaling technique by predicting the power and execution time of a given application over different clock settings. Furthermore, using this frequency scaling by prediction models, we present a deadline-aware application scheduling algorithm to reduce energy consumption while simultaneously meeting their deadlines.

\begin{figure}[t]
\captionsetup{justification=centering}
\begin{subfigure}[t]{0.47\textwidth}
\includegraphics[width=\linewidth]{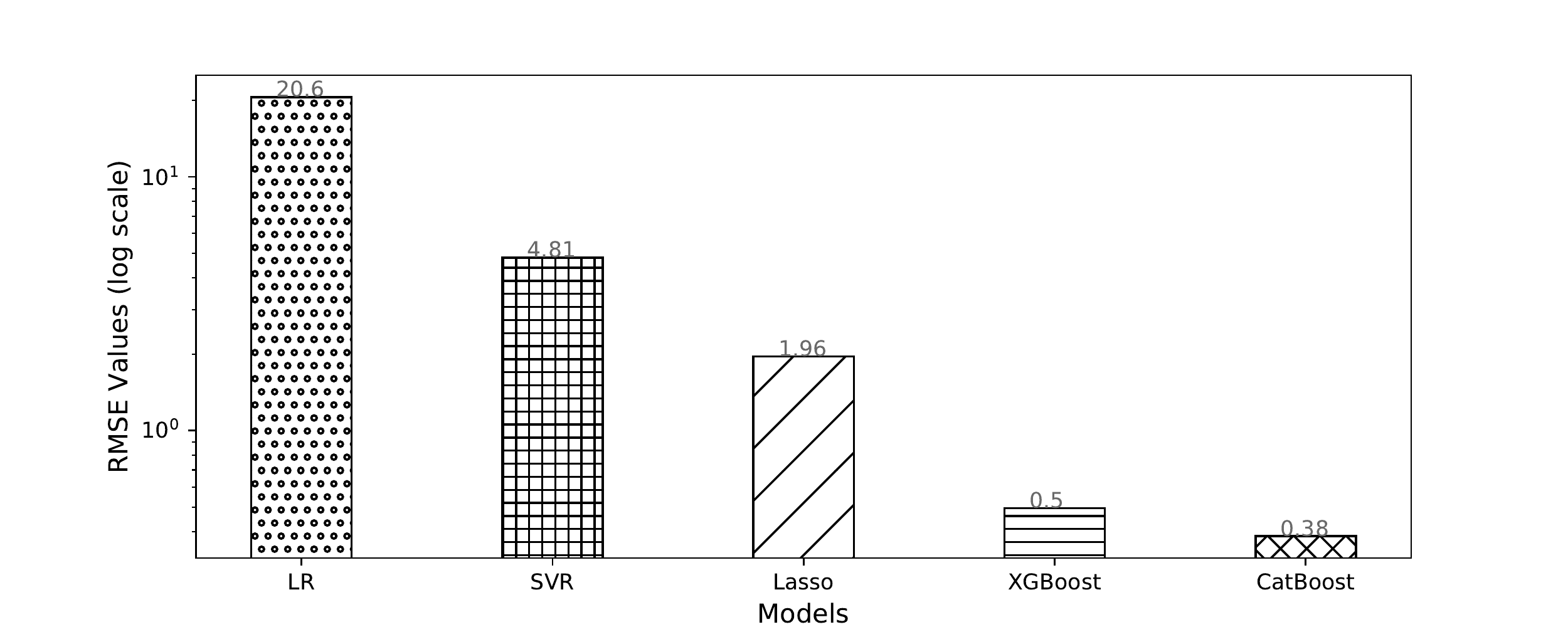}
\caption{Energy prediction}
\label{fig:rmseenergy}
\end{subfigure}
\begin{subfigure}[t]{0.47\textwidth}
\includegraphics[width=\linewidth]{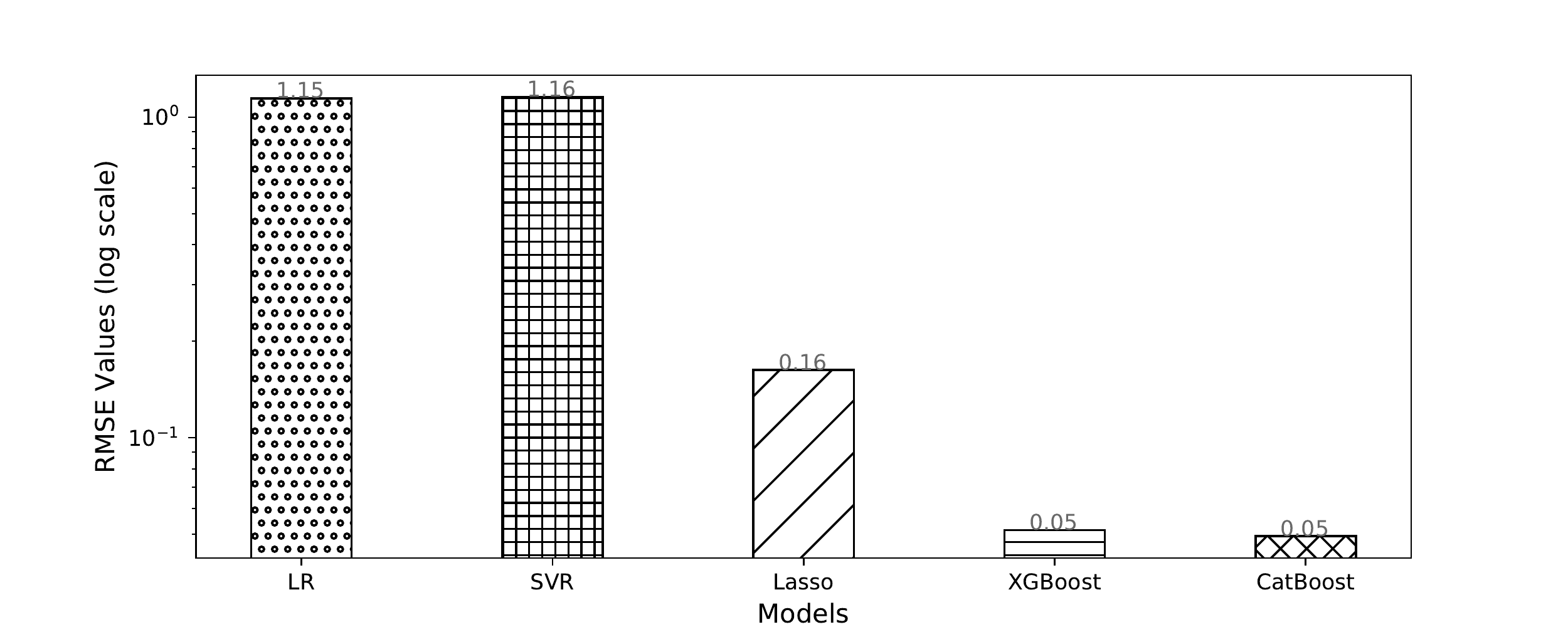}
\caption{Time prediction}
\label{fig:rmsetime}
\end{subfigure}
\caption{Performance of different models for energy and execution time prediction (lower RMSE value is preferred)}
\label{fig:rmse}
\end{figure}

The high-level overview of the system is given in Fig. 3. It is broadly classified into two parts, predictive modelling, and data-driven scheduler. In the first part, we collect the training data that consists of three parts, profiling information, energy-time measurements, and respective frequency configurations. We then predict two entities for a given application and frequency configuration, i.e., energy consumption and execution time. Subsequently, in the second part, the new applications arrive with the dead-line requirements and minimal profiling data from a default clock frequency execution. The scheduler finds correlated application data using the clustering technique, and this data is used for predicting the energy and execution time over all frequencies. Finally, based on the deadline requirements and energy efficiency, the scheduler scales the frequencies and executes the applications.
We use twelve applications for evaluation from two standard GPU bench-marking suites, Rodinia and Polybench. The training data is generated from profiling the applications using nvprof, a standard profiling tool from NVIDIA. We collected around 120 key features representing key architectural, power, and performance counters. To build the predictive models, we explored several regression-based ML models, including Linear Regression (LR), lasso-linear regression (Lasso), and Support Vector Regression (SVR). Also, ensemble-based gradient boosting techniques, extreme Gradient Boosting (XGBoost), and CatBoost. The goal is to build energy and execution time prediction models for each GPU device to assist the frequency configuration.
\begin{figure}
\includegraphics[width=\linewidth]{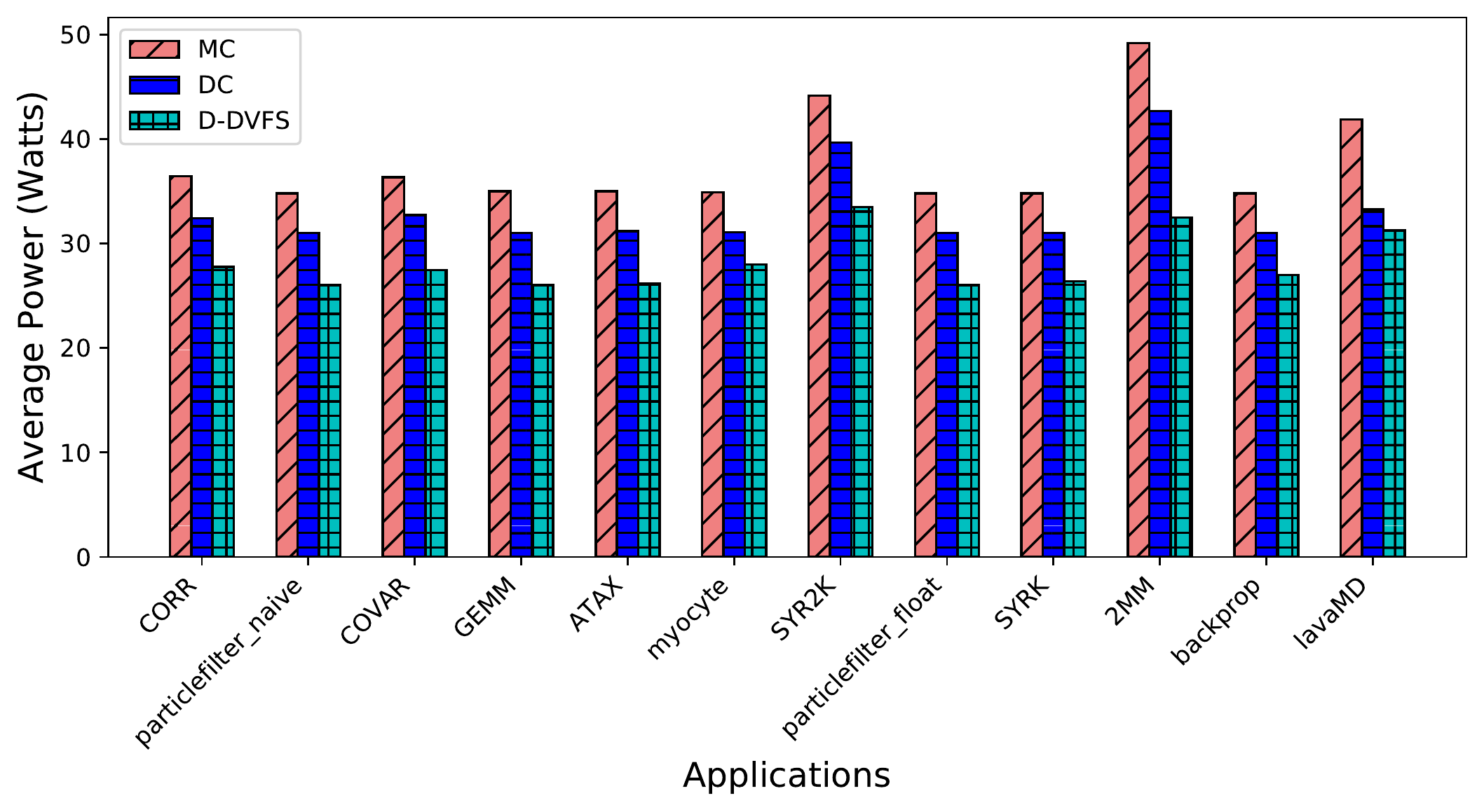}
\caption{Average energy consumption of applications}
\label{fig:application_power}
\end{figure}

\begin{figure}
\includegraphics[width=\linewidth]{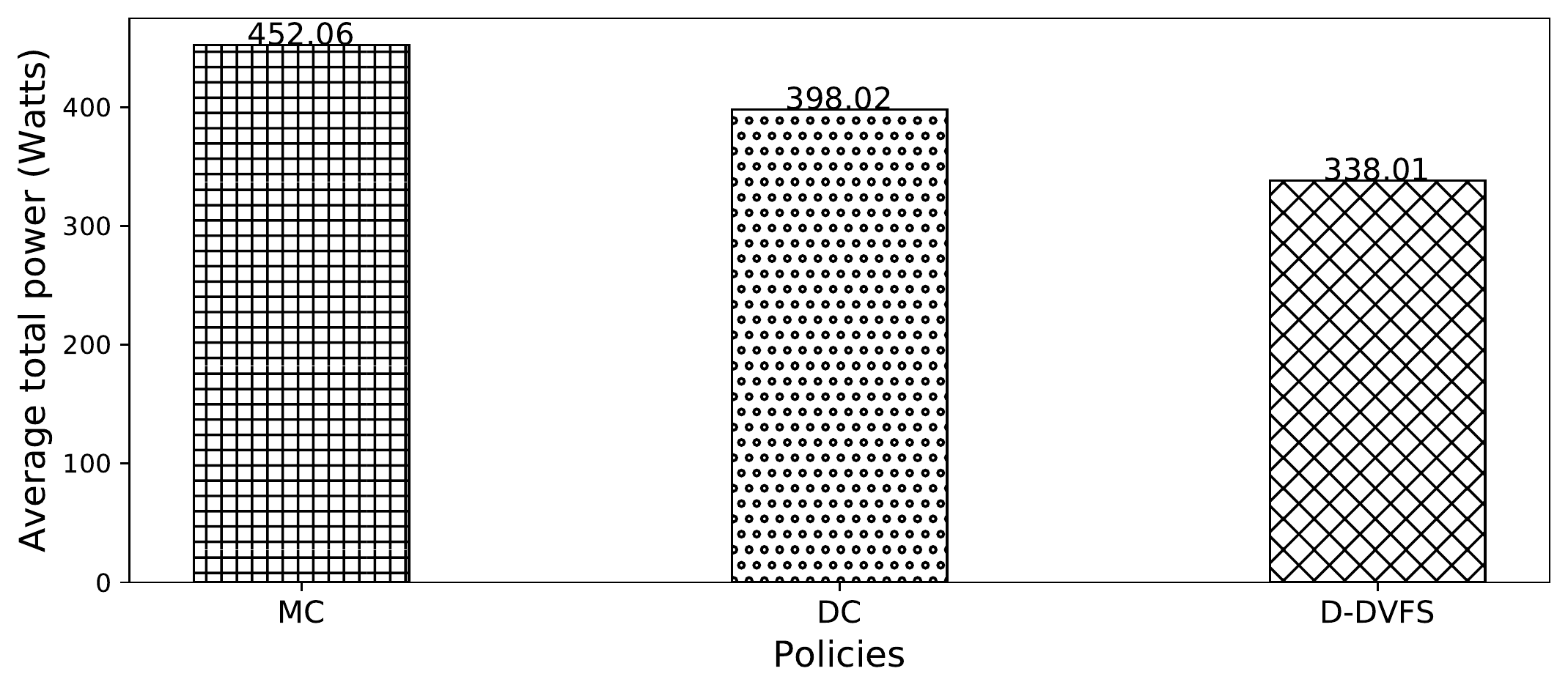}
\caption{Average total energy consumption of GPU} \label{fig:total_power}
\end{figure}

\begin{figure}
\includegraphics[width=\linewidth]{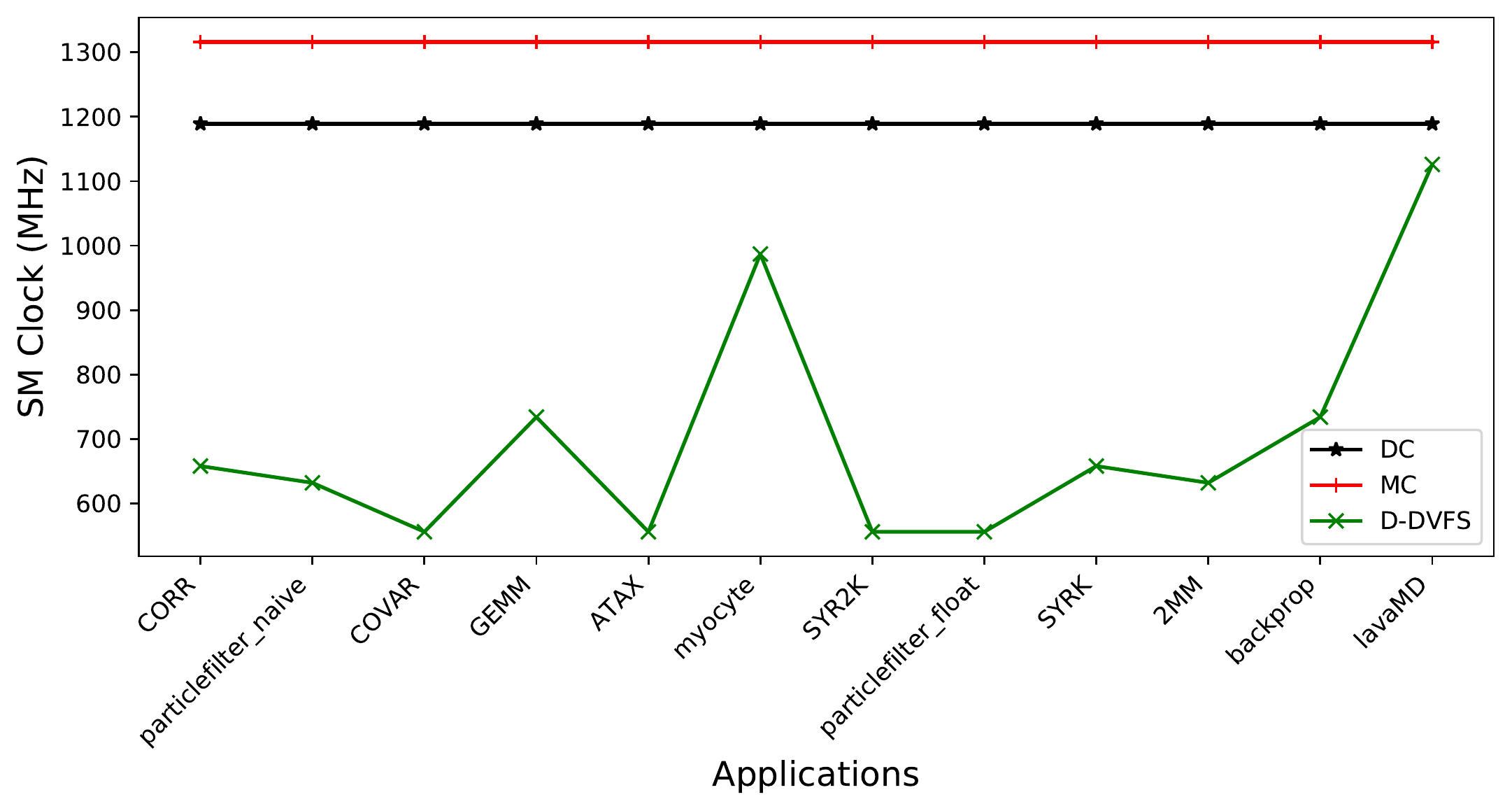}
\caption{Frequency Scaling by different policies}
\label{fig:dvfs_scaling}
\end{figure}

\begin{figure}
\includegraphics[width=\linewidth]{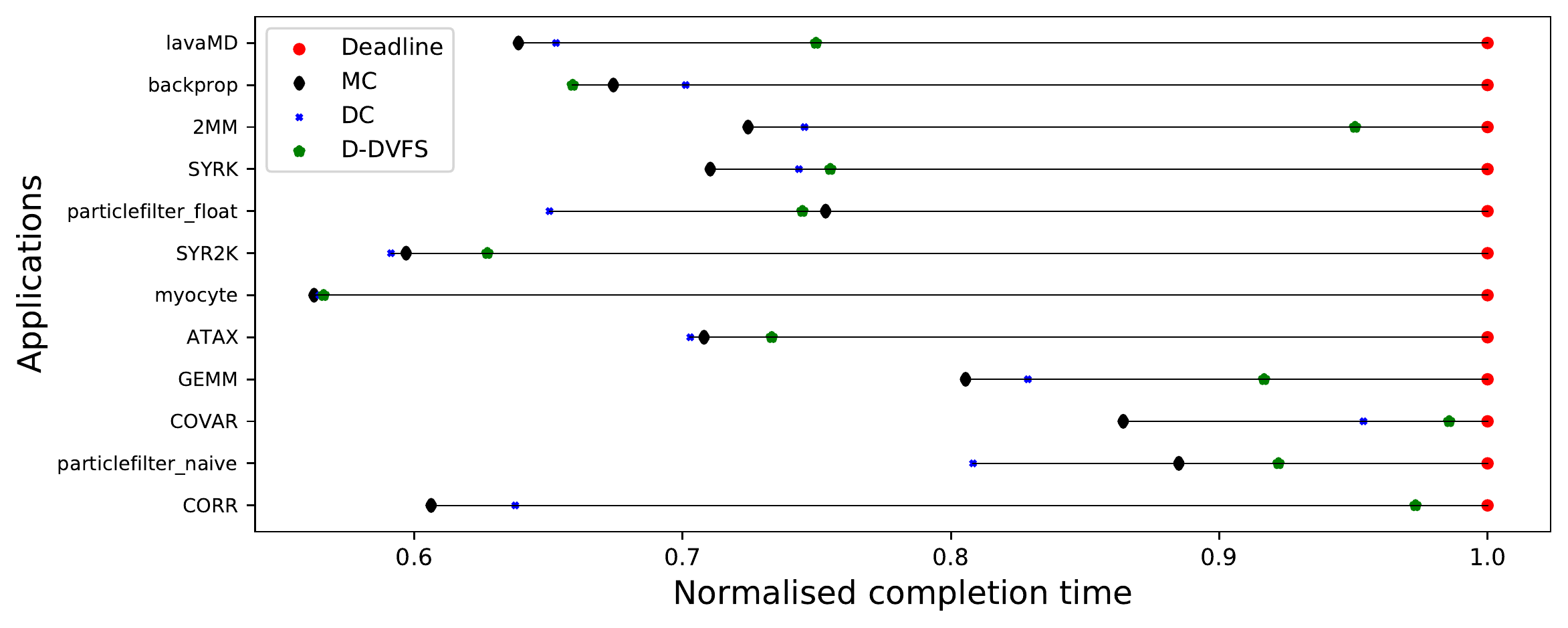}
\caption{Normalised application completion time compared to deadline}
\label{fig:deadline}
\end{figure}

We conduct extensive experiments on NVIDIA GPUs (TESLA P100). The experiment results have shown that our prediction models with CatBoost have high accuracy with the average Root Mean Square Error (RMSE) values of 0.38 and 0.05 for energy and time prediction, respectively (Figure \ref{fig:rmseenergy}, Figure \ref{fig:rmsetime}). Also, the scheduling algorithm consumes 15.07\% less energy (Figure \ref{fig:total_power}) as compared to the baseline policies (default and max clock) while meeting the application deadlines as our approach can scale the frequencies that have energy-efficient settings (Figure \ref{fig:dvfs_scaling}) also able to meet performance requirements. More details on prediction-models, scheduling algorithms, and implementation can be found in \cite{shashiccgrid2020}.
\subsection{Industrial (Google Cloud and Microsoft Azure)  Data Centre Management}
Data centres are the backbone infrastructures of cloud computing today. A data centre is a complex Cyber-Physical-System (CPS) consists of numerous elements. It houses thousands of rack-mounted physical servers, networking equipment, sensors monitoring server, and room temperature, a cooling system to maintain acceptable room temperature, and many facility-related subsystems. The data centre is one of the highest power density CPS of up to 20 kW per rack, dissipating an enormous amount of heat. This poses a serious challenge to manage resources energy efficiently and provide reliable services to users. Optimising data centre operation requires tuning the hundreds of parameters belonging to different subsystems where heuristics or static solutions fail to yield a better result.
Moreover, even a 1\% improvement in data centre efficiency leads to savings in millions of dollars over a year and also helps to reduce the carbon footprints. Therefore, optimising these data centres using potential AI techniques is of great importance. Accordingly, we discuss two real-time AI-based RMS systems built by researchers at Google and Microsoft Azure Cloud.

ML-centric cloud \cite{bianchini2020toward} is an ML-based RMS system at an inception stage from the Microsoft Azure cloud. They built Resource Control (RC)—a general ML and prediction serving system that provides the insights of workload and infrastructure for re-source manager of Azure compute fabric. The input data collected from the virtual machine and physical servers. The models are trained using a gradient boosting tree and trained to predict the different outcomes for user's VMs such as average CPU utilisation, deployment size, lifetime, and blackout time. The Azure resource manager interacts with these models in runtime. For instance, the scheduler queries for virtual machine lifetime, and based on the predicted value; the appropriate decision is taken to increase infrastructure efficiency. Applying these models to several other resource management tasks is under consideration, including power management inside Azure infrastructure.

Similarly, Google has also applied ML techniques to optimise the efficiency of their data centres. Specifically, they have used ML models to change the different knobs of the cooling system, thus saving a significant amount of energy \cite{gao2014machine}. The ML models are built using simple neural networks and trained to improve the PUEs (Pow-er Usage Effectiveness), a standard metric to measure the data centre efficiency. The input features include total IT workload level, network load, parameters affecting the cooling system like outside temperature, wind speed, number of active chillers, and others. The cooling subsystems are configured according to the predictions, and results have shown that the 40\% savings are achieved in terms of energy consumption.
Therefore, the brief uses cases presented here firmly attest to the feasibility of AI-centric solutions in different aspects of resource management of distributed systems.

\section{Conclusions}
Future distributed computing platforms will be massively complex, large scale, and heterogeneous, enabling the development of highly connected resource-intensive business, scientific, and personal applications. Managing resources in such infrastructure require data-driven AI approaches that derive key insights from the data, learn from the environments, and take resource management decisions accordingly. In this paper, we have discussed the challenges associated with the adoption of AI-centric solutions in RMS. We identified the potential future directions describing different RMS tasks where we can apply AI techniques. Moreover, we presented the conceptual AI-centric RMS model. Finally, we demonstrated the two use-cases of AI-Centric solutions in resource management of distributed systems.

The state-of-the-art rule-based or heuristics resource management solutions have become inadequate in modern distributed computing platforms. The RMS policies need to deal with massive scale, heterogeneity, and varying workload requirements. As a result, we believe that AI techniques and tools can be widely utilised in numerous RMS tasks, including monitoring, resource provisioning, scheduling, and many others. Such approaches are highly adaptive and better suited to deal with the resource management complexities, enabling optimised resource management from processor to middleware platforms, and application management.

\bibliographystyle{IEEEtran}

\bibliography{references}

\newpage

\end{document}